\documentclass{emulateapj}

\shorttitle{Variations in the spectral slope of Sgr~A* during a NIR flare}
\shortauthors{Gillessen et al.}

\begin{document}

\title{Variations in the spectral slope of Sgr~A* during a NIR flare}

\author{S.~Gillessen$^1$, F.~Eisenhauer$^1$, E.~Quataert$^2$, R.~Genzel$^{1,\,3}$, 
T.~Paumard$^1$, S.~Trippe$^1$,
T.~Ott$^1$, R.~Abuter$^1$,
A.~Eckart$^4$, P.~O.~Lagage$^5$, M.~D.~Lehnert$^1$, L.~J.~Tacconi$^1$, F.~Martins$^1$}
\affil{$^1\,$Max-Planck-Institut f\"ur extraterrestrische Physik, 85748 Garching, Germany}
\affil{$^2\,$Astronomy Department, University of California, Berkeley, CA 94720, USA}
\affil{$^3\,$Physics Department, University of California, Berkeley, CA 94720, USA}
\affil{$^4\,$1. Physikalisches Institut, Universit\"at zu K\"oln, 50937 K\"oln, Germany}
\affil{$^5\,$UMR 7158, CEA-CNRS-Universit\'e Paris 7, DSM/DAPNIA/Service d'Astrophysique, CEA/Saclay, France}

\begin{abstract}
We have observed a bright flare of Sgr~A* in the near infrared
with the adaptive optics assisted integral field
spectrometer SINFONI\footnote{This work is based on observations
collected at the European Southern Observatory, Paranal, Chile. Program ID: 075.B-0547(B)}. 
Within the uncertainties, the observed spectrum is featureless and can be described
by a power law. Our data suggest that the spectral index  
is correlated with the instantaneous flux and that both
quantities 
experience significant changes within less than one hour.
We argue that the near infrared flares from Sgr~A* are due to synchrotron emission of 
transiently heated electrons, the emission being
affected by orbital dynamics and synchrotron cooling, both
acting on timescales of $\approx 20$ minutes.
\end{abstract}


\section{Introduction}

The detection of stellar orbits \citep{sch02,eis05,ghe03,ghe05} close to Sgr~A* has proven that 
the Galactic Center (GC) hosts a massive black hole (MBH) with a mass of $(3.6\pm 0.2) \times 10^6\,
\mathrm{M}_{\odot}$. Sgr~A* appears rather
dim in all wavelengths, which is 
explained by accretion flow models
\citep{nar95,qua99b,mel00,mel01}.
In the near infrared (NIR) it was detected after diffraction limited
observations at the 8-m class telescopes had become possible \citep{gen03,ghe04}. 
Usually the emission is not detectable. 
However, every few hours Sgr~A* flares in the NIR, reaching up to
$K\approx 15\,$mag. A first flare spectrum was obtained by \cite{eis05},
showing a featureless, red spectrum ($\nu S_{\nu}\sim\nu^\beta$ with
$\beta \approx-3$).

\section{Observations and data reduction}
\label{datared}
We observed the GC on 2005 June 18 from 
2:40 to 7:15 UT with SINFONI \citep{eis03, bon04}, an adaptive optics (AO) assisted integral
field spectrometer which is mounted at the Cassegrain focus
of ESO-VLT Yepun (UT4). 
The field of view was 0.8''$\times$0.8'' for individual exposures, mapped
onto 64$\times$32 spatial pixels. We observed in K-band with
a spectral resolution of FWHM 
$0.5\,$nm. The first 
12 integrations lasted $5\,$min each. During those we noticed 
NIR activity of Sgr~A* and 
we switched to 4 minute exposures. We followed Sgr~A* for another 32 exposures.
In total we interleaved nine integrations on 
a specifically chosen off field (712'' W, 406'' N of Sgr~A*). 
The seeing was $\approx\,$0.5'' and the optical coherence time
$\approx 3\,$ms, some short-time deteriorations excluded. The AO
was locked on the closest optical guide star ($m_R=14.65$, 
10.8'' E, 18.8'' N of Sgr~A*),
yielding a spatial resolution of $\approx 80\,$mas FWHM, close to the
diffraction limit of UT4 in K-band ($\approx 60\,$mas). 

Our detection triggered immediate follow-up observations with VISIR, 
a mid-infrared (MIR) instrument
mounted at ESO-VLT Melipal (UT3). From 5:25 UT onwards 
VISIR was pointing to the
GC. At the position of Sgr~A* no significant flux was seen. A conservative upper limit
of 40$\,$mJy (not dereddened) at 8.59$\,\mu$m is reported (Lagage et al., in prep).

\begin{figure}
\epsscale{1}
\plotone{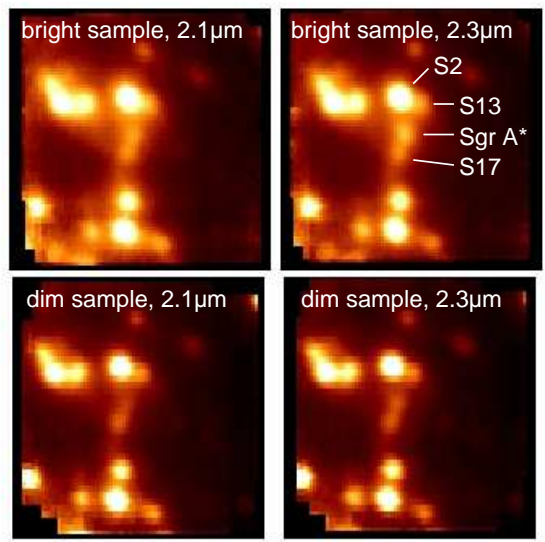}
\caption{The Galactic Center as seen
with SINFONI. Channel maps from
the bright and the dim sample obtained
by taking the mean in spectral dimension
from $2.05\,\mu$m to 
$2.15\,\mu$m and from $2.25\,\mu$m to 
$2.35\,\mu$m. All images are scaled the same way
and use an identical color map.
Sgr~A* is brighter in the
longer wavelength maps, indicating that it is
is redder than the field stars. Furthermore, 
the effect is more pronounced in the dim sample, 
meaning that Sgr~A* is redder therein
than in the bright sample.}  
\label{adb}
\end{figure}

The reduction of the SINFONI data followed the standard 
steps:
From all source data we subtracted the respective sky frames to correct for instrumental
and atmospheric background. We applied 
flatfielding, bad pixel correction, a search for cosmic ray hits,
and a correction for the optical distortions of SINFONI. 
We calibrated the wavelength dimension with line emission lamps and tuned
on the atmospheric OH-lines of the raw frames.
Finally we assembled the data into cubes
with a spatial grid of 12.5$\,$mas/pix.

\section{Analysis}

\begin{figure}
\epsscale{.80}
\plotone{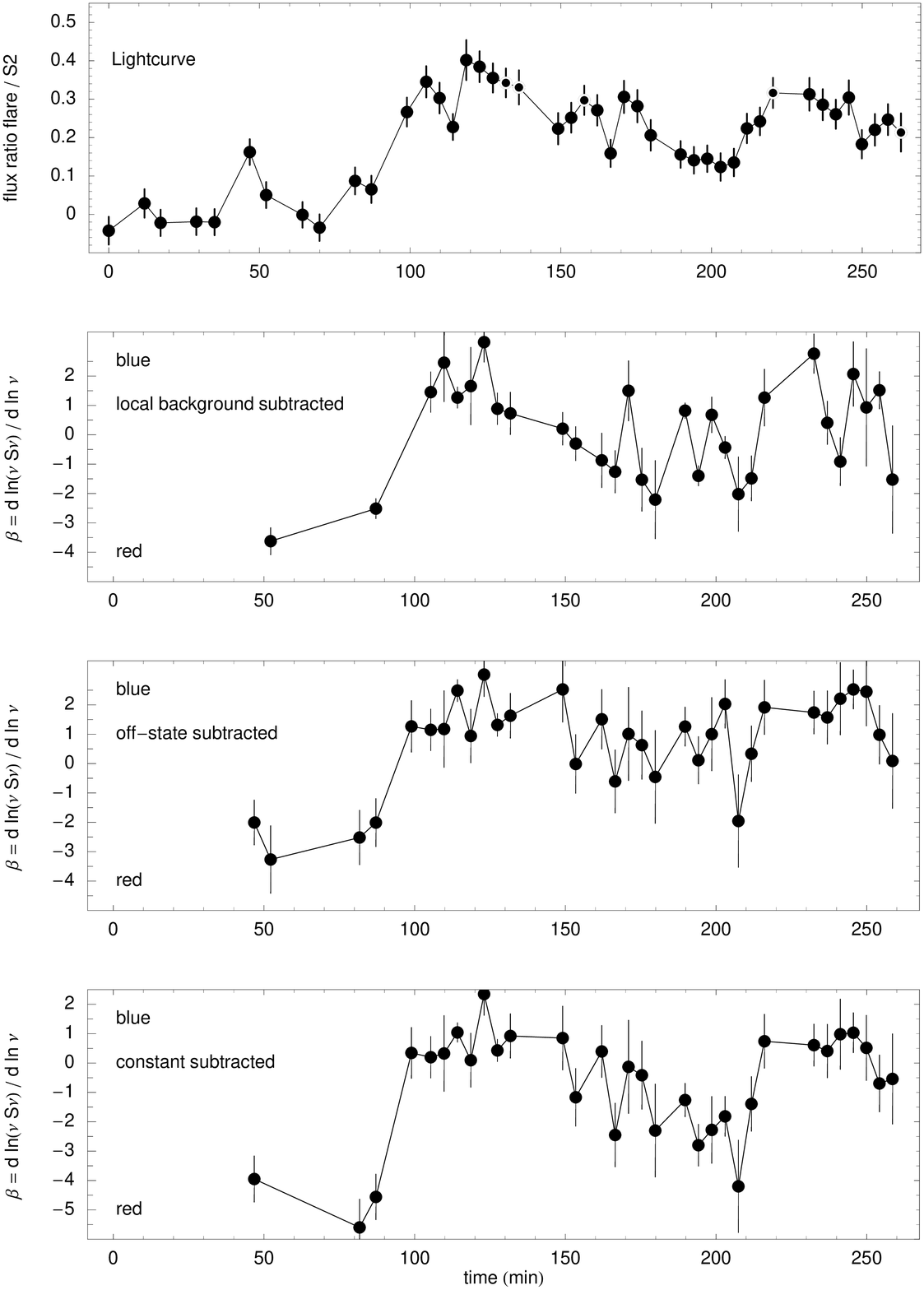}
\caption{Lightcurve and variation of the spectral power law index 
$\beta$ during the flare. 
Time is counted from 2:40 UT. 
Top: Flux ratio flare/S2.
Thin dots are exposures
affected by bad seeing.
Middle/top: $\beta$ using the small apertures background.
Middle/bottom: $\beta$ using the off state subtraction background.
Bottom: $\beta$ using the constant subtraction background.}
\label{lightcurve}
\end{figure}

\subsection{Flux determination}

For all 44 cubes we extracted a 
collapsed image (median in spectral dimension) of 
a rectangular region (0.25''$\times$0.5'') centered on Sgr~A*
and containing the three S-stars S2, S13, and S17.
We determined the flux of Sgr~A*
from a fit with five Gaussians to each of these images. Four
Gaussians with a common width describe the four sources. 
The fifth Gaussian (with a width $3.5\times$ wider, typical for the halo 
from the imperfect AO correction) accounts
for the halo of the brightest star S2 ($K \approx 14\,$mag). The halos of the weaker
sources (all $K < 15\,$mag) could be neglected for the flux measurement. 
We fixed the positions of all sources 
(known a-priori from a combined cube) and
the amplitude ratios for the stars. Five parameters were left free:
An overall amplitude, the background, 
the width, the flux ratio halo/S2, and $\cal{F}$, the
flux ratio Sgr~A*/S2. This procedure disentangles real variability from
variations in the background, the Strehl ratio, and the seeing.

As a crosscheck we determined $\cal{F}$ in a second way for all images;
for both Sgr~A* and S2 we measured the flux difference
between a signal region centered on source and a reasonable,
symmetric background region. 
The such determined values agreed very well
with the fits. For further analysis we
used the fitted ratios and included the difference
between the two estimates in the errors.

\subsection{Choice of background}

The value of the spectral power law index $\beta$ 
crucially depends on the background subtraction. Subtracting too much light
would artificially make the signal look redder than it is. Hence, a
reasonable choice for the background region excludes the nearby sources S2 and
S17. Furthermore, the background flux is
varying spatially. Actually Sgr~A* lies close to
a saddle point in the background light distribution, caused by S2 and S17 (see Fig.~\ref{adb}). 
In the East-West direction 
the background has a maximum close to Sgr~A* and in the North-South direction 
a minimum. 
A proper estimate of the background can be achieved in two ways: 
a) working with small enough, symmetric apertures, and b)
subtracting from the signal an off state obtained at the position of Sgr~A* from cubes in 
which no emission is seen. We used the first method as well as two variants 
of the second.\\
{\bf Small apertures:}
The local background at a position $\vec{x}$ can be estimated by averaging over a small, 
symmetric region centered on $\vec{x}$. 
Given the background geometry we have chosen a ring with inner radius $3\,$pix
and outer radius $7\,$pix. The circular symmetry was only broken since we explicitely excluded those pixels
with a distance to S17 and S2 smaller than 3 and 7 pixels respectively.
Unfavorable of this method is that the local background is only approximated, since a 
sufficiently large region has to be declared as signal region.\\
{\bf Off state subtraction:}
The local background can be extracted from cubes in
which no signal is seen. Since the seeing conditions change from
cube to cube, one still has to correct for the varying amount of stray light in the
signal region. We estimate this variation by measuring the difference
spectrum between signal and off cube in a stray light region.  
The latter must not contain any field stars and should be
as far away from the nearby sources as Sgr~A*.
We used two stray light regions to the left and to the 
right of Sgr~A*, between 5 and 10 pixels away.
The disadvantage of this method is that one needs a suitable off state. 
The latter point is critical for our data. 
Even though Sgr~A* has not been detected directly in 
the first three cubes, the light at its position appears 
redder than the local 
background\footnote{Inspecting older (non-flare) SINFONI data 
we found cases similar to the new data and other cases in which the light
was identical to the local background emission. This is consistent with 
the L-band observations by \cite{ghe04,ghe05b}.}.
Assuming that this
light is due to a very dim, red state of Sgr~A*, we would subtract
too much red light and artificially make the flare too blue. 
In this sense, this method yields an upper limit for $\beta$.\\
{\bf Constant subtraction:}
The off state method can be varied to obtain a lower limit for $\beta$.
Assuming that the true off state spectrum has the color of S2, one can
demand that it is flat after division by S2. In our data, the
S2 divided off state
spectrum is rising towards longer wavelengths. Hence, in this third 
method we estimate the background
at blue wavelengths and use that constant as background for
all spectral bins. 

\begin{figure*}
\epsscale{1}
\begin{center}
\includegraphics[width=5.5cm]{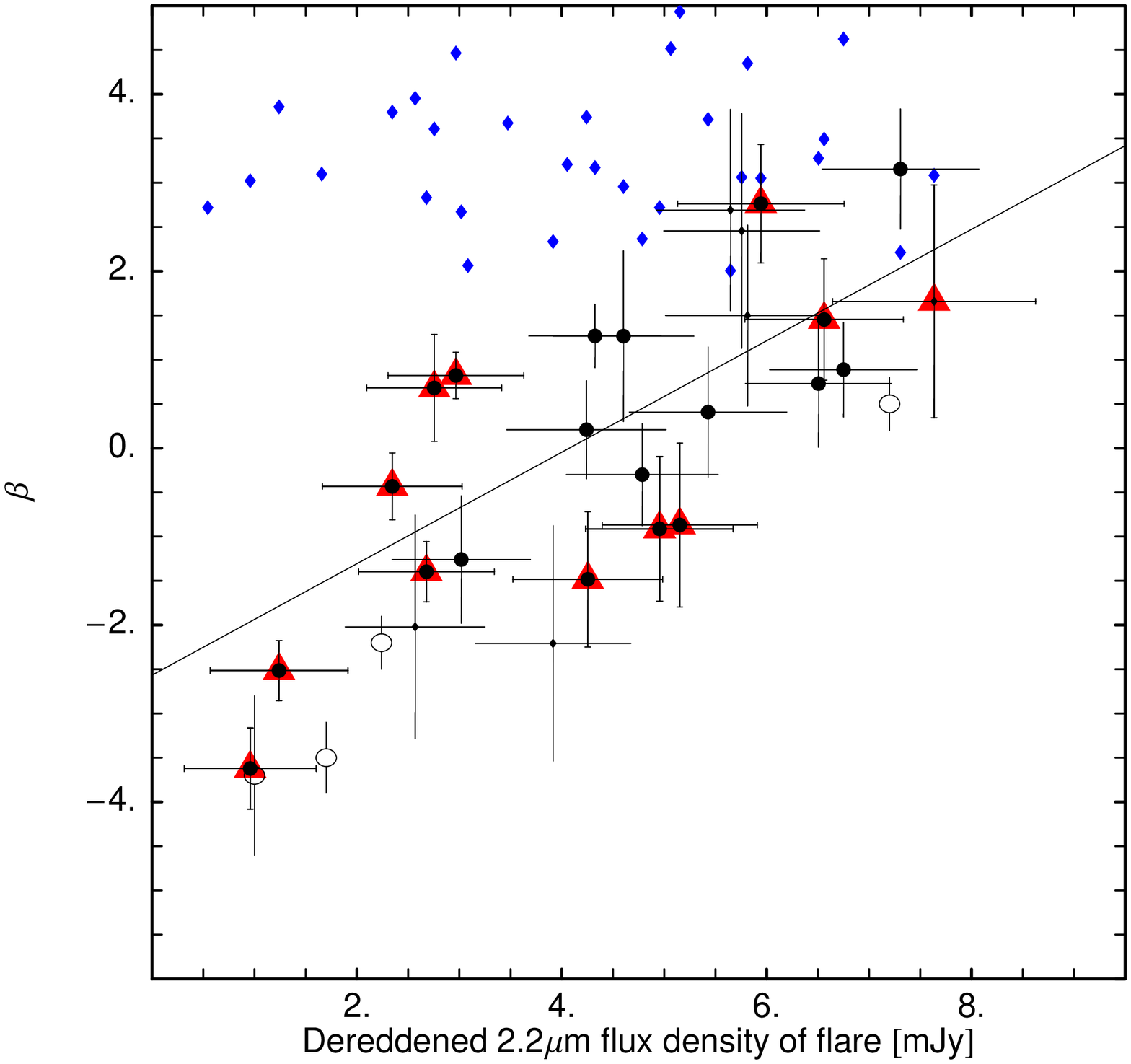}
\includegraphics[width=5.5cm]{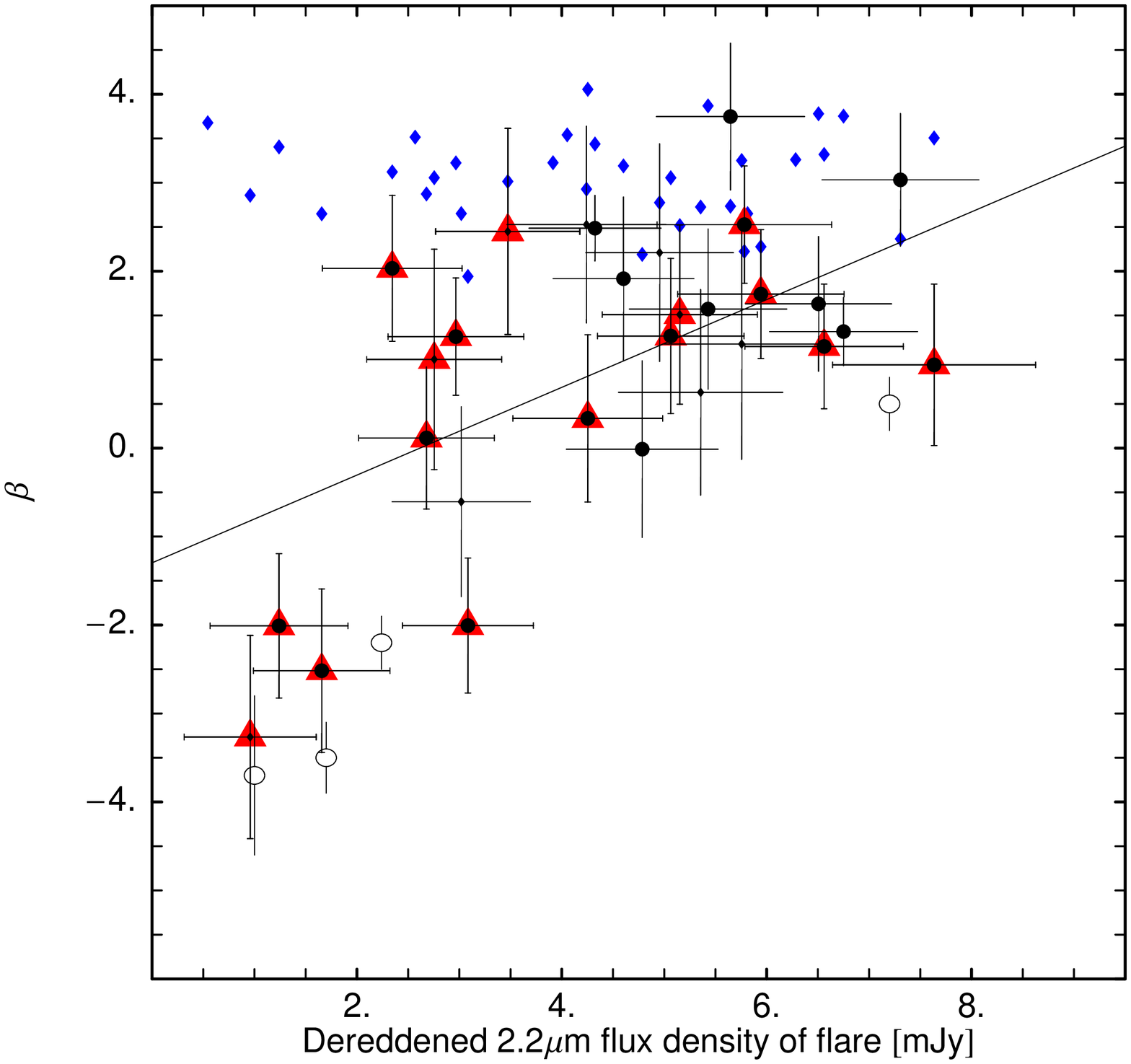}
\includegraphics[width=5.5cm]{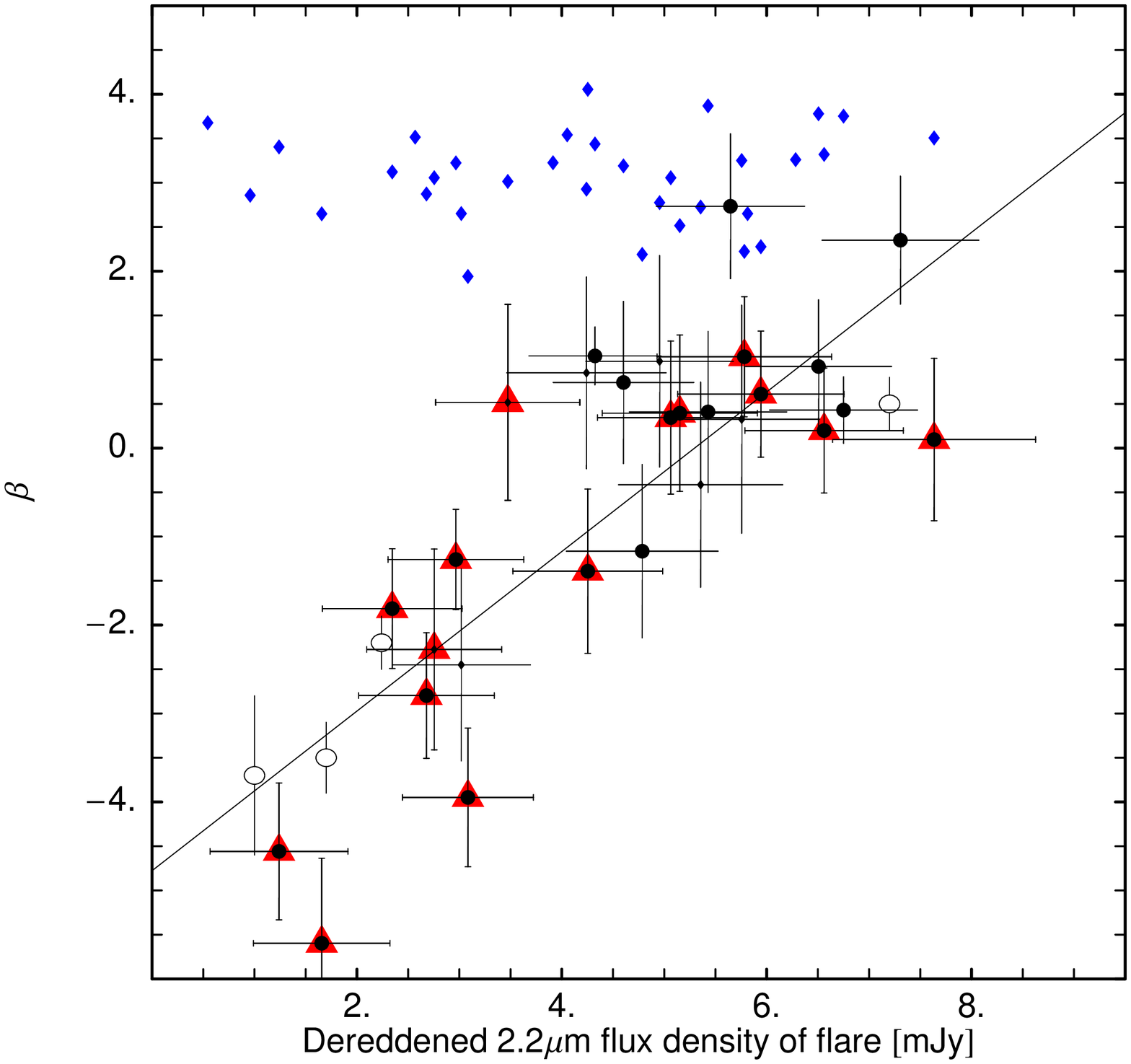}
\end{center}
\caption{Correlation plots between the flare flux and the spectral index $\beta$.
Points with error bars represent the flare, blue dots are $\beta_{S17}$. 
Thick black dots mark the points with an error $\Delta \beta < 1$, red triangles
mark the data with good seeing (FWHM $<75\,$ mas). The line is a fit to the thick black dots.
Open circles denote the data from \cite{eis05} - near 2$\,$mJy - and \cite{ghe05b} 
- near 7$\,$mJy.
Left: Small apertures method; middle: off state subtraction method; right: constant subtraction method.
}
\label{corr}
\end{figure*}

\subsection{Determination of spectral power law index}

We obtained spectra as the median of all pixels inside a 
disk with radius $3\,$pix centered on source minus the median spectrum of the pixels
in the selected off region.
In order to correct for the interstellar extinction 
we then divided by the S2-spectrum (obtained in the same way as the signal). 
Next the temperature of S2 is corrected by multiplying 
by the value of $\nu S_{\nu}$
of a blackbody with $T\approx 25000\,$K. After binning the data
into 60 spectral channels it is finally fit with a power law $\sim \nu^\beta$.
With this definition, red emission has $\beta<0$. The error on $\beta$ 
is obtained as the
square sum of the formal fit error and the standard deviation in a sample of 
20 estimates for $\beta$ obtained by varying the on and off region selection.

\section{Results}
We observed a strong 
(flux density up to $8\,$mJy
or $\nu L_\nu \approx 10^{35}\,$ergs/s), long
(more than 3 hours) flare which 
showed significant brightness variations
on timescales as short as 10 minutes (Fig.~\ref{lightcurve}, top).
While the data is
not optimal for a periodicity search (poorer sampling
than our previous imaging data), it is worth noting that the
highest peak in the periodogram (significance of $\approx 2\,\sigma$) 
lies at a period of $\approx 18\,$min.
This is also the timescale found by \cite{gen03}, who identify the
quasi-periodicity with the orbital time close to the
last stable orbit (LSO) of the MBH.

We divided the data into three groups: a) the cubes of the
first peak near $t=50\,$min ("preflare"), b) the cubes at $t>100\,$min
with $\cal{F}\,$$<$ 0.25 ("dim state"), and c) the cubes at $t>100\,$min
with $\cal{F}\,$$>$ 0.25 ("bright state"). For the three sets
we created 
combined cubes in which we determined  
$\beta$, using
all three background estimates.
In all cases we obtained the correct
spectral index $2.9 \pm 0.5$ for S17 (a star with a spectrum similar
to S2 but a flux comparable to Sgr~A*). For Sgr~A* we get:\\

\begin{tabular}{r|ccc}
$\beta\,\,$&preflare&dim state&bright state\\
\hline
off state subtr.& $-1.4\pm 0.4$ & $-0.7\pm 0.4$ & $+0.4\pm 0.2$\\
small apertures& $-1.8\pm 0.3$ & $-1.7\pm 0.4$ & $-0.1\pm 0.3$ \\
constant subtr.& $-3.4\pm 0.4$ & $-2.3\pm 0.3$ & $-0.3\pm 0.2$
\end{tabular} \\

The values obtained from the small apertures lie
between the values from the other two methods, consistent
with the idea that they yield upper and lower limits.
The absolute values vary systematically according to the chosen background
method. However, independent from that, it is clear that the preflare is
redder than the dim state which in turn is redder than the bright state.
An obvious question then is whether 
flux and $\beta$ are directly correlated.

Hence, we applied the spectral analysis to the individual cubes. 
We kept the data in which Sgr~A* is detected,
the error $\Delta \beta < 1.5$ and the spectral
index for S17 does not deviate more than $1.5$ from
the expected value. The resulting
spectral indices appear to be correlated with the flux 
(Fig.~\ref{lightcurve}, \ref{corr}).
The values match the results in \cite{eis05}
and \cite{ghe05b}.
Bright flares are indeed bluer than weak flares, as suspected
by \cite{ghe05b} and consistent with the earlier
multi-band observations of \cite{gen03}. Our key new result is that 
this even holds
within a single event. 

For all three background methods 
it is clear that the main event was preceded by a weak, red event. 
For the small apertures and the
constant subtraction method instantaneous flux and
spectral index are correlated. For the off state subtraction method one could instead group the
data into a preflare at the beginning and a bluer, brighter 
main event.  

We checked our data for contamination effects. If stray light 
would affect the flare signal, $\beta$ 
should be correlated with the seeing (measured by the width
from the multiple fits). Since we did not find such a correlation,
we exclude significant contamination.

\section{Interpretation}

Theoretical models predict that the mm-IR emission from Sgr~A* is
synchrotron emission from relativistic electrons close to the LSO 
\citep{liu01,qua03,liu04,liu06}.
Radiatively inefficient accretion
flow (RIAF) models with a thermal electron population ($T_\mathrm{e}\approx 10^{11}\,$K)
produce the observed peak in the submm but fail to produce
enough flux at 2$\,\mu$m. The
NIR emission requires transiently heated or
accelerated electrons as proposed by \cite{mar01}.

A conservative interpretation of our data is suggested by
Fig.~\ref{corr}, middle. There was a weak, red event before 
and possibly independent from the
main flare which was a much bluer event.
Plausibly the preflare  
is then due
to the high-energetic tail of the submm peak \citep{gen03}. The main flare
requires nevertheless a population of heated electrons.

A more progressive interpretation follows from
the correlation between flux and $\beta$ (Fig.~\ref{corr}, left \&
right). It suggests that
the NIR variability is caused by the combination of
transient heating with subsequent cooling and orbital dynamics of relativistic
electrons. In the following subsections we will exploit this idea.

\subsection{Synchrotron emission}

In the absence of continued energy injection, synchrotron cooling will suppress the 
high energy part of the electron distribution function. This results in a 
strong cutoff in
the NIR spectrum with a cutoff frequency decreasing with time. At a fixed
band one expects that the light becomes redder as the flux decreases.
This can cause the correlation between
flux and $\beta$. The observed flare 
apparently needs several heating and cooling events.

The synchrotron cooling time is comparable to the observed timescale
of decaying flanks as in Fig.~\ref{lightcurve} (top) or \cite{gen03}.
In a RIAF model \citep{yua03} the magnetic field $B$ is related to the
accretion rate $\dot{M}$. For disk models with
$\dot{M} \approx 10^{-8}\,M_{\odot}/$yr 
\citep{ago00,qua00} one has $B\approx 30\,$G at a 
radial distance of $3.5\,R_\mathrm{S}$ (Schwarzschild radii).
The cooling time for electrons emitting at $\lambda = 2 \,
\lambda_2 \,\mu$m for $B = 30 \, B_{30} \,$G is
\begin{equation}
t_{\rm cool} \approx 8 \, B_{30}^{-3/2} \, \lambda_2^{1/2} \, {\rm min} \,\,. 
\label{cool}
\end{equation}
This is similar to the orbital timescale, making
it difficult to disentangle flux
variations due to heating and cooling from
dynamical effects due to the orbital motion.

\subsection{Orbital dynamics}
\label{dyn}

The timescale of $\approx 20$ minutes for the observed variations 
suggests orbital motion close to the LSO as one possible 
cause. Any radiation produced propagates through strongly curved space-time. 
Beaming, multiple images, and Doppler shifts
have to be considered \citep{hol97,hol99}. 
Recent progress has been made in simulating these effects
when observing a spatially limited emission region orbiting the MBH
\citep{pau05, bro05a, bro05b, bro05c}. 

That the emission region is small compared to the accretion disk
can be deduced from X-ray observations.
\cite{eck04,eck05} report similar timescales for X-ray- and IR-flares.
This excludes that the X-ray emission is
synchrotron light as one
would expect cooling times $<1\,$min
(eq.~\ref{cool}). The X-ray emission is naturally explained by IC scattering of
the transiently heated electrons that have a
$\gamma$-factor of $\gamma\approx 10^3$. 
The IC process scatters the seed photon field up according to
$\nu_\mathrm{IC}\approx \gamma^2 \,\nu_\mathrm{seed}$. Given 
$\nu_\mathrm{IC} \approx 10^{18}\,$Hz for X-rays 
the seed photon frequency is $\nu_\mathrm{seed}\approx 10^{12}\,$Hz - the
submm regime. The X-ray intensity for the synchrotron-self-Compton case is
given by 
\begin{equation}
X/IR = L_\mathrm{SSC}/L_\mathrm{sync} \approx
10^{-2} L_{35} \left(R/R_\mathrm{S}\right)^{-2} B_{30}^{-2}\,\,,
\label{l37}
\end{equation}
where $L_{35}$ is the seed photon field luminosity in units of $10^{35}\,$ergs/s
and $R$ the radius of the emission region. 
The highest value of $L_{35}$ compatible with the MIR- and
submm-data is reached for a flat spectrum ($\beta=0$) of the
heated electrons. Thus,
$L_{35} \lesssim 1$. Via equation~(\ref{l37}) we have 
$R/R_\mathrm{S} \approx 0.1 \sqrt{L_{35}/(X/IR)}$. 
From the observed ratios $X/IR \approx [0.1\,-\,1]$ we derive
$R < 0.3\,R_\mathrm{S}$. That means, the emission region
has to be a small spot. 

A plausible dynamical model
is discussed in Paumard et al. (in prep). 
It reproduces typical lightcurves 
as observed by \cite{gen03} or in Fig.~\ref{lightcurve}.
Due to the Doppler effect
the observed light corresponds to different rest frame
frequencies depending on the orbital phase. If the source spectrum
is curved, 
flux and spectral index appear correlated. Following this
interpretation, 
the emission during the brightest part 
originates from a rest-frequency with larger $\beta$
than the dimmer state emitted at shorter
wavelengths. Such a concavely curved spectrum is
naturally expected from the synchrotron models.

\end{document}